\documentclass[12pt,a4paper]{article}
\usepackage{amsmath}
\usepackage[dvips]{graphicx}
\usepackage{times}
\begin{document}

\begin{titlepage}
\title{Reflective scattering from unitarity saturation}
\author{S.M. Troshin,
 N.E. Tyurin
 \\[1ex]
\small  \it Institute for High Energy Physics,\\
\small  \it Protvino, Moscow Region, 142281, Russia}
\normalsize
\date{}
\maketitle

\begin{abstract}
Proceeding from optical analogy we propose a new physical
interpretation of unitarity saturation  leading to antishadowing
as a reflective scattering. This interpretation of antishadowing
is related to the non-perturbative aspects of strong interactions
and follows from the  specific property of the unitarity
saturation when elastic $S$-matrix $S(s,b)|_{b=0}\to -1$ at $s\to
\infty$. The analogy with Berry phase  and experimental
consequences of the proposed interpretation as reflective
scattering  at the LHC and in the cosmic rays studies are discussed.
\end{abstract}
\end{titlepage}

\section*{Introduction}

 The fundamental
problems of the nonperturbtive QCD  are related to confinement and
spontaneous chiral symmetry breaking phenomena. Those phenomena
deal with collective, coherent interactions of quarks and gluons,
and results in formation of the asymptotic states, which are the
colorless, experimentally observable particles.
 Among the  problems of strong interactions,  the total
cross--section  growth with energy constitutes one of the
important questions.  The nature of this   energy dependence  is
not completely understood since underlying microscopic mechanism
is due to the nonperturbative QCD. An
essential role here belongs to elastic scattering where all hadron
constituents interact coherently and therefore elastic scattering
can also serve as an important tool to the confinement studies.
Owing to the experimental efforts during recent decades it has
became evident that the coherent processes
 survive at  high energies.

General principles  play a guiding  role in the soft hadron interaction studies,
 and, in particular, unitarity which regulates the
relative strength of elastic and inelastic processes is  the most
significant one. It is important to note here that unitarity is formulated
for the asymptotic colorless hadron on-mass shell states and is not directly connected
to the fundamental fields of QCD --- quarks and gluons. The same is valid for
the analyticity which is relevant for the  scattering amplitudes of
the observable particles only.
As it was noted in \cite{doksh}, it is not clear what these fundamental principles
 imply for the confined objects.

Even the extension of $S$--matrix and  $s$--channel unitarity to the
off --mass shell  particle scattering leads to the significant changes
in the predictions for the behavior of the observables: it
 does not rule out now the
power--like asymptotical behavior of the total cross--sections, i.e. unitarity in this case
  does not lead
to the Froissart-Martin bound in the small $x$-region \cite{lopz}.

Our goal here is to provide a new physical interpretation
 for the scattering mode based
on the saturation of the unitarity relation for the on--mass shell particles, when
elastic scattering not only survives, but prevails at
super high energies in hadron collisions at small impact parameter values.
We discuss in this note  this rather
 unexpected behavior  on the base of optical analogy, and list
 some of the respective experimental signatures in the
studies of hadronic interactions at the LHC and in cosmic rays.
It is important to note that some of these predictions are due
to this new interpretation of unitarity saturation as a reason
for the appearance of the reflective scattering at superhigh
energies.

\section{Saturation of unitarity and energy evolution of the geometric
scattering picture}
Unitarity or conservation of probability, which can be   written in terms
of the scattering matrix as
\begin{equation}\label{ss}
SS^+=1,
\end{equation} implies an
existence at high energies of the two scattering modes - shadowing and antishadowing.
Existence of antishadowing has been known long ago \cite{bbla,sachbla},  in the
context of the rising total cross-sections and transition beyond the black disk limit
it was discussed in the framework of the rational unitarization scheme in \cite{phl}.
The most important feature
of antishadowing is self-damping \cite{bbla} of  the
inelastic channels contribution.
Corresponding geometric picture and physical effects will
 be discussed further.
As it was noted in Introduction, the operator $S$ is defined in the Hilbert space
 spanned over vectors
corresponding to the observable physical particles and the unitarity is formulated in terms
of the physical particles.

The attempts to construct
 $S$-matrix in perturbative QCD \cite{ianku}
have difficulties  related to the problem of confinement \cite{hebe,aste}
 and the problem of the undefined
scale dependence entering $S$-matrix through $\alpha_s$ \cite{petrr}. Moreover, as it was shown in
\cite{hanss}, the color fields can be represented by harmonic oscillators (gluons)
only in the limit when the strong interaction coupling constant tends to zero, i.e. at
$Q^2\to \infty$.

After these necessary remarks, one can proceed further with unitarity condition,
apply the standard procedure and derive from Eq. (\ref{ss})
 unitarity relation for the partial wave amplitudes $f_l(s)$ :
\begin{equation}\label{ul}
\mbox{Im} f_l(s)=|f_l(s)|^2+\eta_l(s),
\end{equation}
where elastic $S$-matrix is related to the amplitude as
\begin{equation}\label{sl}
S_l(s)=1+2if_l(s)
\end{equation}
 and $\eta_l(s)$ stands for the contribution of the intermediate inelastic channels to the
elastic scattering with the orbital angular momentum $l$.
The relation (\ref{ul}) turns out to be a quadratic equation in the case of the pure imaginary
scattering amplitude.
 Therefore, the elastic amplitude  appears  to be  not a
singe-valued function of $\eta_l$. But, only one of the two solutions of the equation corresponding
to the relation (\ref{ul})
 is considered almost everywhere
\begin{equation}\label{us}
f_l(s)=\frac{i}{2}(1-\sqrt{1-4\eta_l(s)}),\quad \mbox{i.e.} \quad |f_l|\leq 1/2,
\end{equation}
while another one
\begin{equation}\label{uas}
f_l(s)=\frac{i}{2}(1+\sqrt{1-4\eta_l(s)}),\quad \mbox{i.e.} \quad 1/2 \leq |f_l|\leq 1
\end{equation}
is ignored. Eq.(\ref{us}) corresponds to the typical shadow picture when
elastic amplitude $f_l(s)$ for each value of $l$ is determined by the contribution
of the inelastic channels $\eta_l(s)$ and for small values of the function
$\eta_l(s)$: $f_l(s)\simeq i\eta_l(s)$. Note, that the value $|f_l|=1/2$
corresponds to the black disk limit where
absorption is maximal, i.e. $\eta_l(s)=1/4$. Unitarity limit for the partial wave
amplitude  is unity, i.e. twice as much as the black disk
limit. The saturation of the unitarity limit is provided by Eq.(\ref{uas}); it leads at
the small values of
$\eta_l(s)$ to the  relation $f_l(s)\simeq i(1-\eta_l(s))$.  It means that
elastic amplitude is increasing when contribution of the inelastic channels is decreasing.
Therefore, the term of antishadowing was used.

It is important to consider  the arguments in favor  of the solution
(\ref{uas}) neglecting. They are simple:
it is well known that analytical properties in the complex $t$-plane lead
 to decrease with $l$ at large values of
  $l>L(s)$\footnote{It should be noted that  $L(s)\sim \sqrt{s}\ln s$.}
 of both the amplitude $f_l(s)$ and the contribution of
inelastic channels $\eta_l(s)$ at least exponentially, i.e.
$\lim _{l\to\infty}f_l(s) =0$ and $\lim _{l\to\infty}\eta_l(s) =0$. It is evident that the
 solution (\ref{uas})
does not fulfill the requirement of simultaneous vanishing  $f_l(s)$ and $\eta_l(s)$ at $l\to\infty$
 and this
is the reason for its neglecting. But the requirement of simultaneous vanishing  $f_l(s)$
and $\eta_l(s)$
effective at large values of $l$ only  and, respectively, the solution (\ref{uas}) should be neglected
at large values of $l$. At small and moderate values of $l$ both solutions (\ref{us}) and (\ref{uas})
have equal rights to exist.

But how  both the above mentioned  solutions  can be reconciled and
 realized in the uniform way
  and what physics picture can underlie
 the solution (\ref{uas}), which anticipates saturation of the unitarity limit for the
 partial wave amplitude? In what follows we consider the case and its consequences.
To provide a geometric meaning to the scattering picture we will use an impact
parameter representation.
The unitarity relation
written for the elastic scattering amplitude $f(s,b)$ is
similar to (\ref{ul}) in the high
energy limit and has the following form
\begin{equation}
\mbox{Im} f(s,b)=|f(s,b)|^2+\eta(s,b). \label{ub}
\end{equation}

There is no  universally  accepted
method to implement unitarity
in high energy scattering \cite{sachbla}. The  recent
approaches to solve the problem of high-energy limit in QCD based, e. g. on the AdS scattering,
apriori suppose the eikonal function corresponding  to the black disk \cite{kang}.
The reduced unitarity limit for the amplitude based on Eq. (\ref{us}) is used also in the models
of gluon saturation (cf. \cite{gsat} and references therein).

In principle, a choice of particular unitarization scheme is not completely ambiguous.
The two  above mentioned solutions of unitarity equation (\ref{us}) and (\ref{uas})
can be naturally reconciled and uniformly reproduced by the rational ($U$--matrix) form
of unitarization. The arguments based on analytical properties of the scattering
amplitude  were put forward \cite{blan} in favor of this form.
In the $U$--matrix approach
 the elastic scattering matrix in the
impact parameter representation
is the following linear fractional transform:
\begin{equation}
S(s,b)=\frac{1+iU(s,b)}{1-iU(s,b)}. \label{um}
\end{equation}
 $U(s,b)$ is the generalized reaction matrix, which is considered to be an
input dynamical quantity. The transform (\ref{um}) is one-to-one and easily
invertible.
Inelastic overlap function $\eta(s,b)$
can also be expressed through the function $U(s,b)$ by the relation
\begin{equation}
\eta(s,b)=\frac{\mbox{Im} U(s,b)}{|1-iU(s,b)|^{2}}\label{uf},
\end{equation}
and the only condition to obey unitarity in the form of Eq. (\ref{ub})
 is $\mbox{Im} U(s,b)\geq 0$.

Another way to warrant  unitarity is to represent elastic  $S$-matrix in the
exponential form:
\begin{equation}\label{eik}
S(s,b)=\exp {[2i\delta(s,b)]},
\end{equation}
where $\delta(s,b)$ is a phase shift [$\delta (s,b)\equiv \delta_R (s,b)+i\delta_I (s,b)$]
 and the inequality $ \delta_I(s,b)\geq 0$  is needed to satisfy
unitarity.
Both representations (\ref{um}) and (\ref{eik}) provide $|S(s,b)|\leq 1$
but contrary to (\ref{um}),  $S\neq 0$
for any finite value of $\delta(s,b)$ when $S$  is written in the form  (\ref{eik}).
  To trace a further difference between
them, let us consider for simplicity
 the case of pure imaginary $U$-matrix and make the replacement $U\to iU$.
  The $S$-matrix has the following form
\begin{equation}
S(s,b)=\frac{1-U(s,b)}{1+U(s,b)}. \label{umi}
\end{equation}
 At this point we should make an important remark on the
models, which use rational or exponential forms of the amplitude unitarization.
We note  that most of the models provide increasing
dependence of these functions with energy (e.g. power-like one) and their exponential decrease
with impact parameter $b$. Thus,  any model of this kind for $U$-matrix is not compatible
with any similar eikonal model, predicting crossing the black disk limit.
The value of energy corresponding to this limit
for central collisions  $S(s,b)|_{b=0}=0$
will be denoted as $s_0$ and it is determined by the  equation
$U(s,b)|_{b=0}=1$.
Thus, in the energy region $s\leq s_0$ the scattering in the whole range of impact parameter variation
has a shadow nature and correspond to solution (\ref{us}), the $S$ matrix varies in the range
$0\leq S(s,b)<1$. But when the energy is higher than this threshold
value $s_0$, the scattering picture at small values of impact parameter $b\leq R(s)$
corresponds to the solution Eq. (\ref{uas}), where $R(s)$
is the interaction radius.
The $S$-matrix variation region is $-1<S(s,b)\leq 0$ at $s\geq s_0$
and $b\leq R(s)$. We are going to discuss emerging physical picture of the scattering in this
particular region of impact parameters and very high energies. The schematic energy evolution   of
 $S$-matrix   of the impact parameter dependence is depicted on Fig. 1.
\begin{figure}[hbt]
\begin{center}
\includegraphics[scale=0.5]{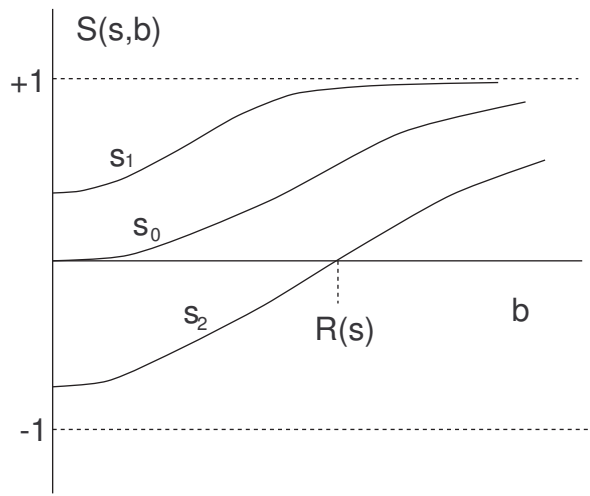}
\caption{\small\it Impact parameter dependence (schematic) of the function $S(s,b)$ for  three values
 of energy $s_1<s_0<s_2$.}
\end{center}
\end{figure}
It is evident that at  $s>s_0$ there is a region of impact parameters where $S$-matrix
is negative,
i.e. phases of incoming state and outgoing  state differ by $\pi$.
Thus, there is a close analogy
here with the light reflection off a dense medium, when the phase of the reflected light is changed by
$180^0$. Therefore, using the optical concepts \cite{gottfr}, the above behavior
 of $S(s,b)$ should be interpreted
as an appearance of a reflecting ability of scatterer due to increase of
 its density beyond some critical value, corresponding to refraction index  noticeably
 greater than unity. In another words, the scatterer has now not only
 absorption ability (due to  presence of inelastic channels), but it starts to be reflective at very
 high energies and its central part ($b=0$) approaches to the completely
 reflecting limit ($S=-1$) at $s\to\infty$.
 It would be natural to expect that this reflection has a diffuse character.
 In another words we can describe emerging physical picture of  very high energy scattering
 as scattering off the reflecting
disk (approaching to complete reflection at the center) which is surrounded by a   black ring.
The reflection which is a result of  antishadowing leads to $S(s,b)|_{b=0}\to -1$ asymptotically.

The inelastic overlap function $\eta(s,b)$ gets a
 peripheral impact parameter dependence in the region $s>s_0$ (Fig.2).
\begin{figure}[hbt]
\begin{center}
\includegraphics[scale=0.75]{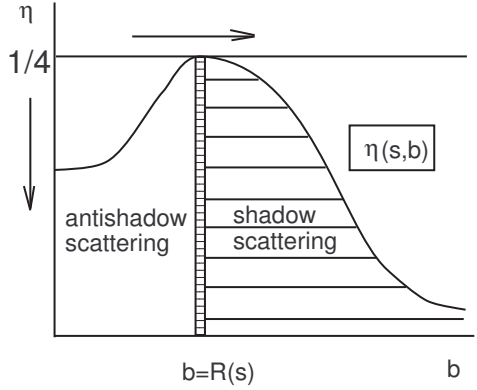}
\caption{\small\it Impact parameter dependence of the inelastic overlap function
 in the unitarization scheme
 with antishadowing. Arrows indicate directions of the energy evolution.}
\end{center}
\end{figure}
Such a dependence is manifestation of the self--damping of the inelastic channels
at small impact parameters.
The function $\eta(s,b)$ reaches its maximum
 value at $b=R(s)$,
 while the elastic scattering (due to reflection) occurs effectively at smaller values
of the impact parameters, i.e.
$\langle b^2 \rangle_{el}<\langle b^2 \rangle_{inel}$. Note that
\[
\langle b^2 \rangle_{i}= \frac{1}{\sigma_i}\int b^2
d\sigma_i \equiv \frac{1}{\sigma_i}\int_0^\infty b^2
\frac{d\sigma_i}{db^2}db^2,
\]
 where $i=tot,el,inel$
and
\[
\mbox{Im} f(s,b)\equiv \frac{1}{4\pi}\frac{d\sigma_{tot}}{db^2};\,\,
|f(s,b)|^2\equiv \frac{1}{4\pi}\frac{d\sigma_{el}}{db^2};\,\,
\eta(s,b)\equiv \frac{1}{4\pi}\frac{d\sigma_{inel}}{db^2}.
\]
The quantity $\langle b^2 \rangle$ is a measure of the particular reaction peripherality and the following
sum rule takes place for $\langle b^2 \rangle_{inel}$:
$\langle b^2 \rangle_{inel}\sigma_{inel}(s)=
\sum_{n\geq 3}\langle b^2 \rangle_{n}\sigma_{n}(s)$.

It is useful to return to the exponential representation for the $S$-matrix (\ref{eik}) at this point.
 It is evident that this form  with pure imaginary phase shift  (eikonal)
 would never give the negative values of the function $S(s,b)$, it will always vary in the
  range $0<S(s,b)<1$. However it is not the case when  $\delta_R (s,b)$
  is not zero. If  $\delta_R (s,b)=\pi/2$, the antishadowing
  can be reproduced in the exponential form of the unitarization, i.e.
  the function $S(s,b)$ will  vary in the
  range $-1<S(s,b)<0$. In order to combine shadowing at large values of $b$ with antishadowing
  in central collisions the real part of the phase shift should have the  dependence
\[
\delta_R (s,b)=\frac{\pi}{2}\theta(R(s)-b).
\]

Such a behavior of $ \delta_R (s,b)$ just takes place in the $U$-matrix form of unitarization.
Indeed, the phase shift $\delta (s,b)$
 can be expressed in terms of the function $U(s,b)$ as following
 \begin{equation}\label{del}
 \delta (s,b)=\frac{1}{2i}\ln\frac{1-U(s,b)}{1+U(s,b)}.
\end{equation}
It is clear that in the region $s>s_0$ the function
 $\delta (s,b)$ has a real part $\pi/2$ in the region $0<b\leq R(s)$, while $\delta_I (s,b)$ goes to
  infinity at $b=R(s)$ (Fig. 3).
\begin{figure}[hbt]
\includegraphics[scale=0.75]{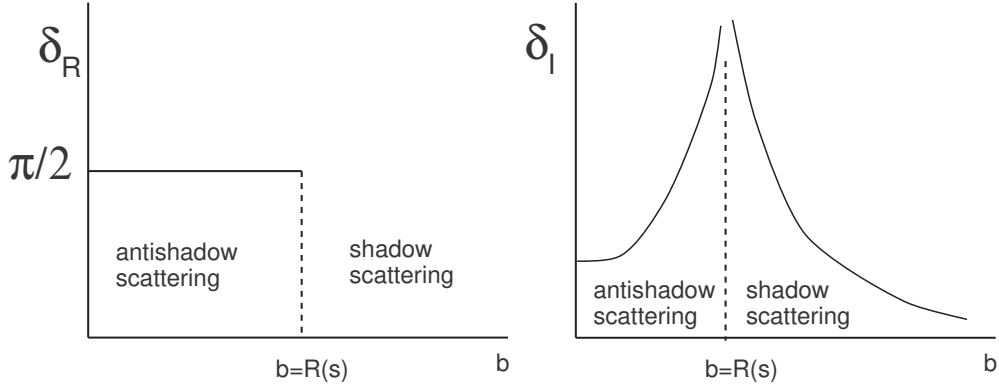}
\caption{\small\it Impact parameter dependence of the real (left panel)
and imaginary part (right panel) of the
 phase shift  at $s>s_0$.}
\end{figure}
The picture leads to discontinuity in the phase shift\footnote{The discontinuity of
$\delta_I (s,b)$ was noticed first by V.A. Petrov \cite{petr}.} and it
resembles  the scattering on the cut-off optical potentials \cite{schre}.

It also leads to another interesting similarity, namely it  allows one to consider $ \delta_R $
 as an analog of the geometric Berry phase in quantum mechanics which appears
 as a result of a cyclic time evolution of the Hamiltonian parameters \cite{berry}.
This interesting phenomenon  can be observed in many physical systems \cite{newton,bphase}
and is, in fact, a feature  of a system that depends only on the path it evolves along of.
 In the case of pure imaginary elastic scattering amplitude the contribution
of the inelastic channels $\eta$ can be considered as a
 parameter which determines  due to unitarity (but not in a unique way)  the elastic $S$-matrix.
 We can  vary variable $s$ (and/or $b$) in a way
 that the parameter $\eta$ (which has a peripheral $b$-dependence, cf. Fig. 2)
  evolves  cyclically from $\eta_i<1/4$ to $\eta_{max}=1/4$ and
  again to the value $\eta_f$, where $\eta_f=\eta_i$ (loop variation).
  As a result
 the  non-zero phase appears ($\delta_R=\pi/2$ at $b\leq R(s)$) and this phase
  is independent of the details
   of the energy evolution.

Thus,  we can summarize  that
the physical scattering picture beyond the black disk limit  evolves with energy
by simultaneous  increase of the reflective ability (i.e. $|S(s,b)|$ increases
with energy and  $\delta_R=\pi/2$),
  and decrease of the absorptive ability $1-|S(s,b)|^2$ at the impact parameters $b<R(s)$.

We address now the elastic scattering amplitude $F(s,t)$, which is a Fourier-Bessel transform
of the $f(s,b)$.  In the $U$-matrix unitarization method
it is determined by the singularities
in the impact parameter complex $\beta=b^2$ plane. Those singularities include
the poles which positions are determined by the solutions of the
equation
$1+U(s,\beta)=0$
and the branching point at $\beta=0$ which follows from the spectral
representation for the function $U(s,\beta)$
\begin{equation}\label{spec}
U(s,\beta)=\frac{\pi^2}{s}\int_{t_0}^\infty \rho(s,t')K_0(\sqrt{t'\beta})dt'.
\end{equation}
With  account of the analytical properties dictated by Eq. (\ref{spec}) we
can  use parameterization  in the form \cite{ttech84,chpr}
\begin{equation}\label{uexp}
 U(s,\beta) =
g(s)\exp(-M\sqrt{\beta^2}/\xi),
\end{equation}
where $M$ is the total mass of $N$  constituent quarks
in the colliding hadrons and $\xi$ is a parameter.
Contribution of
the poles
determines the
 elastic scattering amplitude in the region of small and moderate values of $-t$.  The
amplitude dependence in this region  provides diffraction
 peak and  shows up dip-bump
structure of the differential cross-section. It reproduces also
Orear behavior at larger values of $-t$.

The pole and
cut contributions are decoupled dynamically when $g(s)\rightarrow
\infty  $ at $s\rightarrow \infty$ \cite{ttech84,chpr}.
At large angles  the contribution
from the branching point
is a dominating one. The large angle or small impact parameter scattering
being a result of the reflection
has a power-like angular distribution dependence:
\begin{equation}\label{pow}
\frac{d\sigma}{dt}\propto \left(\frac{1}{s}\right)^{N+3}\omega(\theta).
\end{equation}
The power-like
dependence of the differential cross sections in large angle
scattering closely interrelates with the rise  of the total
cross sections at high energies since both are determined by the
dependence of $g(s)\rightarrow
\infty  $ at $s\rightarrow \infty$ in the
 unitarity saturating scheme.

For the symmetric case of $pp$-interaction the scattering is the same
in the forward and backward hemispheres. The more interesting  case is the one
where interacting particles are not identical, for example,
$\pi N$-scattering. In this case
the behavior of the differential cross--section in the forward hemisphere
 is completely analogous to the above case
of $pp$-scattering where overall behavior of the differential
cross-section  incorporates
diffraction cone, Orear type and power-like dependencies.
 But in the backward hemisphere the poles contributions are suppressed
compared to the cut contribution in the whole region of the variation of the variable $u$ , i.e.
the  power-like dependence will take place at all values of $u$ and there would be no
diffraction cone and Orear type dependence \cite{ttech84} in the backward hemisphere.
It is not surprising, indeed, if we will recollect reflecting nature
of the scattering at small impact parameters.

\section{Observable effects at the LHC energies}
In the $U$--matrix
unitarization scheme  asymptotical behavior of
the cross--sections and  the ratio of elastic to total
cross-section are different from the asymptotic equipartition of elastic and inelastic
cross--sections based in the black disk limit  when
\[
\sigma_{el}(s)/\sigma_{tot}(s)\to\frac{1}{2}.
\]
Under the $U$-matrix unitarization the rise of the elastic cross--section
 is predicted to be steeper than the rise of
the inelastic cross--section beyond some threshold energy. Asymptotically
\begin{equation}\label{umat}
\sigma_{tot}(s)\sim\sigma_{el}(s)\sim\ln^2 s,\quad
\sigma_{inel}(s)\sim\ln s.
\end{equation}
This is due to saturation of the upper unitarity limit for the partial--wave
 amplitude in the $U$--matrix
approach  $|f_l|\leq 1$, while  the restriction  $|f_l|\leq 1/2$
dictated by the black disk limit leads to the above mentioned
 equipartition of the cross-sections.
Note, that despite  the asymptotics for
$\sigma_{el}$ and $\sigma_{inel}$ are different, the quantities $\langle b^2 \rangle_{el}$ and $
\langle b^2 \rangle_{inel}$  have the same energy dependence, proportional to $\ln^2 s$ at
$s\to\infty$.

The above asymptotic dependencies take place for various forms of the function
$U(s,b)$. Explicit forms\footnote{Of course, quantitative predictions are different for the
different models while qualitative features coincide.} for the function $U(s,b)$
 can be obtained, e.g. using geometrical,  Regge
or chiral quark models for the $U$--matrix.
Of course, it is useful to have numerical estimates for the cross-sections at the LHC
energies. These estimates are model dependent ones and can vary rather
strongly depending on the choice of the particular model parameterizations for $U$-matrix.
However, both the Donnachie-Landshoff
and dipole Pomeron parameterizations of $U$-matrix, used in \cite{lasl} and the model
\cite{chpr} are in agreement that the black disk limit will be passed at $\sqrt{s}=2$ TeV.
 The latter model
provides the following values  at the LHC energy $\sqrt{s}= 14$ TeV:
$\sigma_{tot}\simeq 230$ mb
and
$\sigma_{el}/\sigma_{tot}\simeq 0.67$ \cite{pras}.
Thus, there is an interesting possibility that the reflective scattering mode could be discovered
at the LHC by measuring $\sigma_{el}/\sigma_{tot}$ ratio which would be
greater than the black disk  value $1/2$.
However,
the asymptotical regime (\ref{umat})
is expected in the model at $\sqrt{s}> 100$ $TeV$ only.

Proceeding from an increasing weight at the LHC energies of the  reflective scattering
compared to the shadow scattering, we can suppose that the dip-bump structure in the $d\sigma/dt$
in $pp$ scattering might be less prominent at large values of $-t$ and hadronic glory
effect (enhancement of backscattering probability) would be observed.
Unfortunately, it is difficult to give a more definite
predictions for differential cross-sections, since  other effects
 such as real part of the amplitude and/or contributions from helicity-flip amplitudes
  should be taken into account.

The listed above  values for the global characteristics of
$pp$ -- interactions at the LHC  are different from the
other model predictions. First, the total cross--section is predicted
to be twice as much  the common predictions in the range 95-120
mb \cite{vels,bl} due to strong increase of the $\sigma_{el}(s)$.
The prediction for the inelastic cross-section is
$\sigma_{inel}(s)=75$ mb
and coincides with the predictions of the other models \cite{bl}. Thus, reflective scattering would not
result in  worsening  the background situation at the LHC, since the elastic scattering provides
only two extra particles in the final state.
 However, the prediction for the total cross-section   overshoots even
  the existing cosmic ray data. But
 extracting the total proton--proton cross sections from cosmic ray
 experiments is   far from being straightforward
 (cf. e.g. \cite{vels,bl} and references therein). Indeed, those experiments
  are sensitive to the model
 dependent parameter called inelasticity and they do not
 the elastic channel. If we will use for the ratio
 $\sigma_{el}/\sigma_{tot}$ the value $0.67$ at the LHC energy $\sqrt{s}=14$ TeV
 and recalculate the total cross-section obtained
 from the cosmic ray data we will get the value about 227 mb in good agreement with the
 predicted value. This agreement, however, is merely an indirect confirmation of the model prediction
  for the total cross-section, since as it was noted there is an ambiguity in the elastic cross-section
  determination. The behavior of the ratio
  ${\sigma_{el}}/{\sigma_{tot}}$ at
  $s\to\infty$ does not imply decreasing energy dependence of $\sigma_{inel}$.
  The inelastic cross--section $\sigma_{inel}$ increases monotonically
  and  grows as $\ln s$ at $s\to\infty$. Such a dependence of $\sigma_{inel}$ is
  in good agreement with the experimental data and, in particular, with the observed
   falling slope of the depth of shower maximum distribution \cite{gaisser}.
We will discuss some of the cosmic ray related
 issues in the next section.
\section{Reflective scattering
 in cosmic rays}
As it was already noted, the important role in the studies of cosmic rays belongs to
 the  inelasticity parameter $K$,
which is defined as ratio of the energy going to inelastic processes to the total energy.
 The energy dependence of $K$ is not
 evident and cannot be directly measured. The number of models predict
  its decreasing energy dependence while other
 models insist on the increasing energy behavior at high energies \cite{shabel}.
 Adopting simple ansatz of geometric models where parameter
 of inelasticity
 is related to inelastic overlap function we can use the following equation \cite{dias}
 \[
 \langle K \rangle=4\frac{\sigma_{el}}{\sigma_{tot}}
 \left(1-\frac{\sigma_{el}}{\sigma_{tot}}\right)
\]
to get a qualitative knowledge on the inelasticity energy dependence.

The estimation   based on the particular model with
 antishadowing \cite{pras} leads to increasing dependence  with energy
 till $E\simeq (3-4)\cdot 10^7$ GeV. In this region inelasticity reaches maximum value
 $\langle K \rangle = 1$, since ${\sigma_{el}}/{\sigma_{tot}}=1/2$ and then
 starts  to decrease at the energies where this ratio goes beyond the black
  disk limit $1/2$.
Such qualitative
 non-monotonous energy dependence of inelasticity is the result of transition to
  the reflective
 scattering regime.

Reflective scattering results in relative suppression of particle
production at small impact parameters:
\begin{equation}\label{mm}
\bar n(s)= \frac{1}{\sigma_{inel}(s)}{\int_0^\infty  \bar n
(s,b)\frac{d\sigma_{inel}}{db^2}db^2}
\end{equation}
 due to the
 peripherality of $ {d\sigma_{inel}}/{db^2}$.
Thus, the main contribution to
the integral  multiplicity $\bar n(s)$ comes from
the region of $b\sim R(s)$  and the distinctive
feature of this mechanism  is the ring-like shape of particle production which will
lead to correlations in the transverse momentum of the secondary particles.
It means that the enhancement of particle production
at fixed impact distances $b\sim R(s)$ would lead to higher probability of the circular  events.
 Such events
would reflect the production geometry with complete absorption at the impact distances equal
to the effective   interaction radius $R(s)$.
The enhancement of the peripheral particle production
  would destroy the balance
 between orbital angular momentum in the initial and final states; most of the particles
 in the final state would carry  out  orbital
 angular momentum.
To compensate this orbital momentum the spins of the secondary particles should  become
  lined up, i.e. the spins of the  particles in the ring-like events should demonstrate
   significant correlations. Of course the observation of such
effects is difficult due to  the
 multiple interaction of the secondary particles
in the atmosphere.
However, the circular event has been observed experimentally \cite{mount}.

It might be useful to note that the rescattering processes
  are affected by the reflective scattering. At the energies and impact parameters
  where reflective scattering exists there will be no rescatterings, i.e. the
head-on collisions will experience a reflective elastic scattering
at the energies $s>s_0$ and due to this fact
less number of secondary
 particles will be detected in EAS at the ground level.  This should provide
a faster decrease of the energy spectrum reconstructed from EAS, i.e. it
  will result in the appearance of the knee.
 Thus,   the hadron interaction and mechanism of particle
generation will be changing in the region of $\sqrt{s}=3-6$ TeV.
Indeed, the energy spectrum which follows a simple
power-like law $\sim E^{-\gamma}$ changes its slope in this energy
region, i.e. index $\gamma$ increases from
$2.7$ to $3.1$.
 The interpretation of the
cosmic-ray data is complicated since the primary energies
 of cosmic particles
are far beyond of the energies of  modern accelerators with fixed targets and
existing simulation programs merely extrapolate the present knowledge on the hadron
interaction dynamics in the unexplored energy  region. This might be an oversimplification
and here we would like to interpret
the existence of the knee in the cosmic ray energy spectrum
as the effect of  changing hadron interaction mechanism related to the appearance of
the reflective scattering
at small impact parameters.

\section*{Conclusion}
In this note we considered saturation of the unitarity limit in head-on (and small impact
parameter) hadron collisions, i.e. when $S(s,b)|_{b=0}\to -1$ at $s\to\infty$. Approach
 to the full
absorption in head-on collisions,
 in another words, the limit $S(s,b)|_{b=0}\to 0$ at $s\to\infty$ does not follow
from unitarity, it is merely  a result of the assumed
saturation of the black disk limit. This limit is a direct consequence of the exponential
unitarization with an extra assumption on the pure imaginary nature of the phase shift.
On the other hand, the reflective scattering
is a natural interpretation of the unitarity saturation  based on the optical concepts in
high energy hadron scattering.
 Such reflective scattering can be interpreted as a result of the
continuous  increasing density of
the scatterer with energy, i.e. when density goes beyond some critical value relevant
for the black disk limit saturation,
 the scatterer starts to
acquire  a reflective ability.
Having in mind the quark-gluon hadron structure it would be tempting to find the particular
 microscopic  mechanisms  related
 to  the collective quark-gluon dynamics in head-on collisions which can be envisaged as the
 origin of the reflection phenomenon.
 One can try to speculate at this point and  relate the appearance of the reflective scattering
  to the Color-Glass Condensate in QCD (cf. e.g. \cite{mcler} and references therein) merely ascribing
  the reflective ability to Glazma.
However, as it was mentioned in the Introduction,
  the evident obstacle which preclude
  a direct link to the microscopic mechanisms is the problem of confinement.
  The unitarity and consequently
its saturation and reflective scattering are the concepts relevant for the hadronic degrees of freedom.
The concept of reflective scattering itself is  general,
 and results from the $S$-matrix  unitarity saturation
related to  the necessity to provide an indefinite total cross section growth at $s\to\infty$.

Thus, at very high energies ($s>s_0$)   two separate  regions of
 impact parameter distances can be anticipated, namely the outer region
of peripheral collisions where the scattering has a typical shadow origin, i.e.
$S(s,b)|_{b>R(s)}>0$ and
 the inner region of central collisions
where the scattering has a combined reflective and absorptive origin, $S(s,b)|_{b< R(s)}<0$.
 The transition to the negative
values of $S$ leads to
the appearance of the real part of the phase shift, i.e. $\delta_R(s,b)|_{b< R(s)}=\pi/2$.

The generic geometric
 picture at fixed energy beyond the black disc limit can be described
 as a scattering off
the partially reflective and partially absorptive disk
surrounded by the black ring which becomes grey at larger values of the
impact parameter. The evolution with energy  is characterized
by increasing albedo due to the  interrelated  increase of reflection
  and decrease of absorption at small impact parameters. This picture implies
  that the scattering amplitude at
  the LHC energies is beyond the black disk limit at small impact
  parameters (elastic $S$-matrix is negative)
  and it  provides  explanation for the regularities
  observed in cosmic rays studies.
\section*{Acknowledgement}
We are grateful to  V.A. Petrov  for
the interesting discussions.

\end{document}